\RequirePackage{fix-cm}

\documentclass[smallextended]{svjour3}       
\smartqed  

\usepackage{graphicx}
\usepackage{amsfonts}
\usepackage{color}
\usepackage{soul}
\journalname{Quantum Studies: Mathematics and Foundations}

\begin{document}

\title{
Partial Transposition in a Finite-Dimensional Hilbert Space:
Physical Interpretation, Measurement of Observables and Entanglement  
}


\titlerunning{Partial Transposition, Measurement, and Entanglement}        

\author{Yehuda B. Band      
\and 
Pier A. Mello 
}

\authorrunning{Band and Mello} 

\institute{Y. B. Band \at
Department of Chemistry, Department of Physics, Department of Electro-Optics, and the Ilse
Katz Center for Nano-Science,
Ben-Gurion University, Beer Sheva, Israel, 84105, and\\
New York University and the NYU-ECNU Institute of Physics at NYU Shanghai, 3663 Zhongshan Road North, Shanghai, 200062, China \\      
            \email{band@bgu.ac.il}           
           \and
           Pier A. Mello \at
              Instituto de F\'isica, Universidad Naciional Aut\'onoma de M\'exico, 
              Apartado Postal 2-364, M\'exico D. F., 01000 Mexico \\
            \email{mello@fisica.unam.mx}
              }

\date{Received: date / Accepted: date}

\maketitle


\begin{abstract}

We show that partial transposition for pure and mixed two-particle states in a discrete $N$-dimensional Hilbert space is equivalent to a change in sign of  a ``momentum-like" variable of one of the particles in the Wigner function for the state.  This generalizes a result obtained for continuous-variable systems to the discrete-variable system case.  We show that, in principle, quantum mechanics allows measuring the expectation value of an observable in a partially transposed state, in spite of the fact that the latter may not be a physical state.  We illustrate this result with the example of an ``isotropic state", which is dependent on a parameter $r$, and an operator whose variance becomes negative for the partially transposed state for certain values of $r$; for such $r$, the original states are entangled.


\end{abstract}

\section{Introduction}  
\label{intro}

Quantum entanglement in multipartite qubit (and qunits, i.e., $N$-dimensional quantum bits) states is a powerful computation and information resource \cite{Steane_98,band-avishai}.  Entanglement of pure quantum states is well understood, but entanglement of mixed quantum states, i.e., states that cannot be represented using a wave function but must be described using a density matrix, is not yet fully understood.  For pure bipartite states, Schmidt coefficients relate the degree of entanglement to the von Neumann entropy of the reduced density matrix associated with either of the two subsystems; a pure state with a reduced density matrix possessing a vanishing von Neumann entropy corresponds to a separable state, whereas one with finite von Neumann entropy is entangled, and one with maximum von Neumann entropy is maximally entangled.  But no general measure of entanglement of mixed states exists \cite{Horo_01}.  Even deciding whether a state is entangled or not is not always an easy task for mixed states.  A large variety of measures have been studied in the literature to quantify entanglement for a given state, as discussed in Ref.~\cite{plenio_virmani_2006}.  Entanglement witnesses, i.e., functionals which can determine whether a specific state is separable or not, have been proposed \cite{plenio_virmani_2006}.  A useful concept in this context is {\em partial transposition} (PT) with respect to one of the particles \cite{peres_96,horodecki_97,horodecki_98}: when the partially transposed state is not a legitimate quantum mechanical (QM) state, the original state is entangled.

It has been noted in Ref.~\cite{simon_2000} (see also Refs.~\cite{werner-wolf2001,braunstein_van_loock_2005}) 
that, for continuous variables, partial transposition of one particle of a bipartite state amounts to a change in sign of the momentum of that particle in the Wigner function (WF) of the state. 
For the case of discrete variables one can define a ``coordinate-like" and a ``momentum-like" variable 
\cite{mello_revzen_2014,mann_mello_revzen_2016}. 
Here we prove that for the discrete variables case, PT can be interpreted in terms of a change in sign of a momentum-like variable of one of the particles in the Wigner function of the state.  
Just as for the continuous case, this statement is appealing, as it gives an intuitive interpretation of PT.
For this purpose, the generalization of the concept of Wigner function to the discrete-variable case is needed.
This generalization has been widely studied
(see, e.g., Ref.~\cite{mello_revzen_2014} and references cited therein); 
here we use the formulation developed in Refs.~\cite{mello_revzen_2014,mann_mello_revzen_2016}.  

A PT ``state" may not be a physically realizable state.  However, given an observable $\hat{A}$ and a state $\hat{\rho}$, we show that there exists a Hermitian operator $\hat{A}^{T_1}$ with the property that the expectation value of $\hat{A}$ in the PT state $\hat{\rho}^{T_1}$ is the same as the expectation value of $\hat{A}^{T_1}$ in the original, bona-fide state $\hat{\rho}$.  Thus, in principle, the determination of $\langle \hat{A} \rangle_{\hat{\rho}^{T_1}}$ using the original state $\hat{\rho}$ is allowed by Quantum Mechanics.

We discuss positive-definite operators with respect to a bona-fide state $\hat{\rho}$, but having a negative expectation value in the PT state $\hat{\rho}^{T_1}$; this signals entanglement in the original state $\hat{\rho}$.  According to the statement of the previous paragraph, the expectation value of the corresponding PT operators with respect to $\hat{\rho}$ is negative.  This is the meaning of expressions like ``negative variance" that we shall use frequently in the paper.

We illustrate these results with the example of an 
``isotropic state", which is a mixed state constructed as a convex combination of a Bell state $|\Phi^{+} \rangle$ and the completely incoherent state, i.e., $\hat{\rho}_r = r |\Phi^{+} \rangle \langle \Phi^{+}| + (1-r)\frac{\hat{\mathbb{I}}}{N^2}$, where $r$ is a real parameter.  We find that for the PT $\hat{\rho}_r^{T_1}$ , the variance of certain operators becomes negative for $r_0 < r < 1$, thus signaling entanglement of the original state; $r_0$ is obtained as a function of the dimensionality $N$.  From the theorem mentioned in the above paragraph, this variance is in principle measurable, so that entanglement of the original state is detectable.

This paper is organized as follows.  In Sec.~\ref{schwinger} we introduce the Schwinger operators, discuss partial transposition of a state of a bipartite system and show how momentum and position operators for a finite-dimensional system can be defined (subsections \ref{schwinger-1-part} and \ref{schwinger-2-parts} specifically treat one-particle and bipartite systems).  Section \ref{measurement in PT1} shows how one can measure an observable in a partially transposed state (even when such a state is not a physically realizable state).  In Sec.~\ref{consequences for positive operator- observable A} we explore the consequences of the theorem for obtaining the expectation value of an observable in a partially transposed state introduced in the previous section, and in Sec.~\ref{consequences for positive operator-Omega arbitrary} we discuss the consequences of the theorem for a positive-definite operator expressed in terms of an arbitrary operator.  Section \ref{illustration_general_Omega} provides an example for a positive-definite operator of arbitrary dimension $N$ using the isotropic state for arbitrary Hilbert space dimensionality, and finally, Sec.~\ref{conclusion} provides a summary and conclusion.  
Appendices \ref{schwinger-one-particle}, \ref{property of P(p1,p2)}, 
\ref{proof of PT and W discrete} and \ref{proof <Omega>-var(Omega)-T1}
provide some further information on Schwinger operators for one-particle states, and prove some results discussed in the main body of the paper.


\section{Schwinger operators, partial transposition and change in sign of the momentum}
\label{schwinger}

By way of introduction, we first consider one-particle, whose description  can be modelled in terms of a discrete $N$-dimensional Hilbert space.  We then extend the analysis to two-particle systems, which is the main topic of this paper.

\subsection{One-particle system}
\label{schwinger-1-part}

Consider a one-particle system with a discrete, finite set of states.  The eigenvalues of observable operators take on a discrete set of values and the quantum description is given in terms of a finite-dimensional Hilbert space.  As an example, consider a system with angular momentum $j$, described in a Hilbert space of dimensionality $2j+1$.  Another example is the position and momentum observables taken on a discrete lattice of finite dimensionality $N$ (see, e.g., Ref.~\cite{de_la_torre-goyeneche}).  The latter case is the one we shall explicitly work with in this paper.

The Hilbert space to be considered is thus spanned by $N$ distinct states $|q\rangle$, with $q=0, 1, \cdots, (N-1)$.  As discussed in 
Appendix~\ref{schwinger-one-particle}, the periodicity condition $|q+N\rangle = |q\rangle$ is imposed.  The Schwinger operators \cite{schwinger} $\hat{X}$ and $\hat{Z}$ are also defined in 
Appendix~\ref{schwinger-one-particle}, as are the operators $\hat{q}$ and $\hat{p}$.  
Because $\hat{X}$ {\em performs translations in the variable $q$ and $\hat{Z}$ 
in the variable $p$, we regard $\hat{q}$ and $\hat{p}$ as ``position-like" and ``momentum-like" operators}, respectively.
Note, however, that their commutation relation for finite $N$ is quite complicated [e.g., see Ref.~\cite{de_la_torre-goyeneche}, Eq.~(20)], and that in the continuous limit their commutator reduces to the standard form 
\cite{de_la_torre-goyeneche,durt_et_al}, $[\hat{q}, \hat{p}] = i$.

Appendix~\ref{property of P(p1,p2)} shows that under transposition of the density matrix in the coordinate representation for $N>2$ (for $N=2$, $| p\rangle = |-p\rangle$; this restriction does not arise if we do not discuss the transformation $p \rightarrow -p$ of the momentum-like variable), the probability distribution of momentum is affected as follows:
\begin{equation}   \label{P(p) 1part}
P_{\rho^{T}}(p) = P_{\rho}(-p).
\end{equation}
Thus, transposition in the {\em coordinate representation} has the intuitive physical meaning  of changing the sign of momentum $p$ in the momentum probability distribution, an effect which corresponds to {\em time-reversal} (if no spin is present).  Moreover, the Wigner function defined in Refs.~\cite{mello_revzen_2014,mann_mello_revzen_2016}
has the property,
\begin{equation}
W_{\hat{\rho}^{T}}(q,p)
=W_{\hat{\rho}}(q,-p) \; ,
\label{PT and W discrete 1part}
\end{equation}
as demonstrated in Appendix~\ref{proof of PT and W discrete}, thus again exhibiting a {\em change in sign of} $p$.  The definition of the Wigner function in Refs.~\cite{mello_revzen_2014,mann_mello_revzen_2016} requires $N$ to be a prime number larger than 2.  It turns out that this is the simplest extension of the continuous case to the discrete one that one can study, which can then be extended to the case where $N$ is not prime (see, e.g., Ref.~\cite{wootters87,wooters-fields89}).  When $N$ is a prime number, the integers $0, \cdots, N-1$ form a mathematical field playing a role parallel to that of the field of real numbers in the continuous case.  Also, in this case a set of $N+1$ mutually unbiased basis states is known \cite{ivanovic36}.  In what follows, when the Wigner function is not involved, the prime dimensionality requirement is not needed.

\subsection{Two-particles}
\label{schwinger-2-parts}

Let us now consider the two-particle case, which is the one of special interest here.  Each particle is described in an $N$-dimensional Hilbert space.   We shall use Schwinger unitary operators defined for each particle and relations similar to Eqs.~(\ref{X(p)}) and (\ref{Z(q)}) of Appendix~\ref{schwinger-one-particle} to introduce the operators $\hat{p}_i$ and $\hat{q}_i$, which play the role of ``momentum-like" and ``position-like" operators for particle $i$.  
Appendix~\ref{property of P(p1,p2)} shows that under partial transposition of particle 1, for $N>2$ 
(we recall that for $N=2$, $| p\rangle = |-p\rangle$), the joint probability distribution of the two momenta is affected as follows:
\begin{equation}  \label{P(p1,p2)}
{\cal P}_{\hat{\rho}^{T_1}}(p_1, p_2)
= {\cal P}_{\hat{\rho}}(-p_1, p_2).
\end{equation}
Thus, $PT_1$ in the coordinate basis has the intuitive physical meaning  of changing the sign of momentum $p_1$ for particle 1 in the joint probability distribution of the two momenta.  The Wigner function, defined as in Refs.~\cite{mello_revzen_2014,mann_mello_revzen_2016}, has the property, shown in Appendix~\ref{proof of PT and W discrete},
\begin{equation}
W_{\hat{\rho}^{T_1}}(q_1,q_2;p_1,p_2)
=W_{\hat{\rho}}(q_1,q_2;-p_1,p_2) \; ,
\label{PT and W discrete}
\end{equation}
thus exhibiting again a {\em change in sign of} $p_1$.  
Recall that the definition of the Wigner function of Refs.~\cite{mello_revzen_2014,mann_mello_revzen_2016} requires $N$ to be a prime number larger than 2 (see discussion in the previous subsection for one particle).

\section{Measuring an observable in a partially transposed ``state"}
\label{measurement in PT1}

It would appear that measuring the expectation value of an observable in a PT state is impossible when the latter is not a physical state.  But, in fact, such a measurement is allowed by quantum mechanics, as we now show.

Consider a Hilbert space of finite dimensionality $N$ and a Hermitian operator $\hat{A}$ defined in it.
Its expectation value in the state $\hat{\rho}$ is
\begin{equation}
\langle \hat{A} \rangle_{\hat{\rho}}
={\mathrm{Tr}}(\hat{\rho} \hat{A})
= \sum_{m_1 m_2, n_1 n_2} \hat{\rho}_{m_1 m_2, n_1 n_2} 
A_{n_1 n_2, m_1 m_2} \; .
\label{<A>rho}
\end{equation}
The expectation value in the PT ``state" $\hat{\rho}^{T_1}$ is
\begin{eqnarray}
\langle \hat{A} \rangle_{\hat{\rho}^{T_1}}
&=& {\mathrm{Tr}}(\hat{\rho}^{T_1} \hat{A})
= \sum_{m_1 m_2, n_1 n_2} 
(\hat{\rho}^{T_1})_{m_1 m_2, n_1 n_2} A_{n_1 n_2, m_1 m_2}    
\nonumber 
\\
&=&  \sum_{m_1 m_2, n_1 n_2} 
\hat{\rho}_{n_1 m_2, m_1 n_2} A_{n_1 n_2, m_1 m_2} ,
\nonumber 
\\
(n_1 \Leftrightarrow  m_1) &=&  \sum_{m_1 m_2, n_1 n_2} 
\hat{\rho}_{m_1 m_2, n_1 n_2} A_{m_1 n_2, n_1 m_2}
\nonumber 
\\
&=& \sum_{m_1 m_2, n_1 n_2} 
\hat{\rho}_{m_1 m_2, n_1 n_2} ({A^{T_1}})_{n_1 n_2, m_1 m_2} 
\nonumber 
\\
&=& {\mathrm{Tr}}(\hat{\rho} \hat{A}^{T_1}) ,
\label{<A>rhoT1 e}
\end{eqnarray}
where
\begin{equation}
(\hat{A}^{T_1})_{n_1 n_2, m_1 m_2}
= A_{m_1 n_2, n_1 m_2}  .
\label{<A>T2}
\end{equation}
Thus 
\begin{equation}
\langle \hat{A} \rangle_{\hat{\rho}^{T_1}}
= \langle \hat{A}^{T_1} \rangle_{\hat{\rho}} \; .
\label{<A>vs<A>T1}
\end{equation}
The operator $\hat{A}^{T_1}$ is Hermitian. Indeed
\begin{eqnarray}
{\rm From \;  Eq.} \;\;\; (\ref{<A>T2}):  
&& \;\; \langle n_1 n_2| \hat{A}^{T_1}  | m_1 m_2  \rangle
= \langle m_1 n_2| \hat{A} | n_1 m_2  \rangle       
= \langle   n_1 m_2  |  \hat{A}^{\dagger} | m_1 n_2  \rangle^{*}
\nonumber      \\
(\hat{A}=\hat{A}^{\dagger}) \Rightarrow   && 
= \langle   n_1 m_2  |  \hat{A} | m_1 n_2  \rangle^{*}  
= \langle   m_1 m_2 |  \hat{A}^{T_1} |n_1 n_2  \rangle ^{*} \; ,
\end{eqnarray}
showing that
\begin{equation}
\hat{A}^{T_1}
= (\hat{A}^{T_1})^{\dagger}.
\label{A=Herm}
\end{equation}

As a result, we have the following:
\newtheorem{statement}{Theorem}
\begin{statement} 
Given an observable $\hat{A}$ and a state $\hat{\rho}$, there exists a Hermitian operator $\hat{A}^{T_1}$ with the property that the expectation value of $\hat{A}$ in the PT ``state" $\hat{\rho}^{T_1}$ has the same value as the expectation value of $\hat{A}^{T_1}$ in the original, bona fide state $\hat{\rho}$ (as opposed to $\hat{\rho}^{T_1}$ which may not be a physically realizable state), and is thus, in principle, amenable to measurement.
\label{theorem}
\end{statement}


\section{Consequences of Theorem \ref{theorem} for obtaining the expectation value of an observable in a partially transposed state}
\label{consequences for positive operator- observable A}

Consider a Hermitian operator $\hat{A}$.
The operator $A^2$ is {\em positive-definite} with respect to the {\em bona fide} state
$\rho$, i.e.,
\begin{equation}
\langle A^2 \rangle_{\hat{\rho}} 
= {\mathrm{Tr}}(\rho A^2) \ge 0.
\label{A2 ge 0}
\end{equation}

On the other hand, $A^2$ may not be positive-definite with respect to
${\rho}^{T_1}$ which, in general, is not a bona fide state and may have negative eigenvalues, i.e.,
\begin{equation}
\langle A^2 \rangle_{\hat{\rho}^{T_1}}
= {\mathrm{Tr}}(\rho^{T_1} A^2) \;\;\; {\rm may \; not \; be} \ge 0 .
\label{A2 nge 0 1}
\end{equation}
Now, Theorem \ref{theorem} applied to Eq.~(\ref{A2 nge 0 1}) gives 
\begin{equation}
\langle A^2 \rangle_{\hat{\rho}^{T_1}}
=\langle (A^2)^{T_1} \rangle_{\hat{\rho}}
\;\;\; {\rm may \; not \; be} \ge 0 .
\label{A2 nge 0 2}
\end{equation}
How can the RHS of (\ref{A2 nge 0 2}) not be $\ge 0$, in spite of $\hat{\rho}$ being a 
bona fide QM state?
The reason is that $(A^2)^{T_1}$ {\em may not be a positive-definite operator}, i.e.,
\begin{equation}
\langle (A^2)^{T_1} \rangle_{\hat{\rho}}
= {\mathrm{Tr}}[\rho (A^2)^{T_1}] 
\stackrel{in \; general}{\neq}
{\mathrm{Tr}}[\rho  (A^{T_1})^2 ] \ge 0 \; , 
\label{A2 nge 0 3}
\end{equation}
due to the fact that 
\begin{equation}
(A^2)^{T_1} 
\stackrel{in \; general}{\neq}
(A^{T_1})^2 \ge 0 \; . 
\label{A2 nge 0 4}
\end{equation}
I.e., 
\begin{eqnarray}
{\rm although} \;\;\; 
&& (A^{T_1})^2 \ge 0 \;\; {\rm with \; respect \; to \; a \; bona \; fide \; \rho},   
\label{AT12 ge 0}      \\
&& (A^2)^{T_1}
\stackrel{may \; not \; be}{\ge 0} \; {\rm with \; respect \; to \; a \; bona \; fide \; \rho} .
\label{A2 nge 0 5}
\end{eqnarray}

As a consequence, if we discover a (Hermitian) observable $\hat{A}$
such that the PT operator $(\hat{A}^2)^{T_1}$ (a Hermitian operator, and thus an observable) 
has a negative expectation value in the state $\hat{\rho}$,
i.e.,
$\langle (\hat{A}^2)^{T_1}\rangle_{\hat{\rho}} < 0$, then $\hat{\rho}$ is entangled.

\subsection{Illustration for $N=2$.}
\label{N=2}

As an example for $N=2$, consider the two-particle entangled pure state
\begin{eqnarray}
| \Phi^{+} \rangle 
&=& \frac{1}{\sqrt{2}}(|00\rangle + | 11 \rangle) ,
\label{Phi+} \\
\hat{\rho}
&=&  | \Phi^{+} \rangle \langle \Phi^{+}|
\label{rho_Phi+},
\end{eqnarray}
and the observable
\begin{equation}  \label{A}
\hat{A} = \sigma_{1x}\sigma_{2x}+\sigma_{1y}\sigma_{2y}+\sigma_{1z}\sigma_{2z},
\end{equation}
which has the property
\begin{equation}  \label{A2}
\hat{A}^2 = 3-2\hat{A} .
\end{equation}
Here, $\sigma_{1\alpha}$ is the Pauli matrix $\sigma_{\alpha}$ for particle 1 and similarly for particle 2,
and $|0\rangle = |\! \uparrow \rangle$ and $|1\rangle = |\! \downarrow \rangle$.
We find the first moment, second moment and variance of $\hat{A}$ to be given by
\begin{eqnarray}
\langle \hat{A} \rangle_{\hat{\rho}}
&=& 1 ,
\label{<A>rho} \\
\langle \hat{A}^2 \rangle_{\hat{\rho}}
&=& 1 ,
\label{<A2>rho} \\
{\mathrm{Var}}(\hat{A})_{\hat{\rho}} &=& 0,
\label{varArho}
\end{eqnarray}
consistent with $| \Phi^{+} \rangle$ being an eigenstate of $\hat{A}$
with eigenvalue 1.
Under partial transposition $T_1$, the various operators transform as
\begin{eqnarray}
A^{T_1}
&=&\sigma_{1x}\sigma_{2x}-\sigma_{1y}\sigma_{2y}+\sigma_{1z}\sigma_{2z}
\label{AT1}  \\
(A^2)^{T_1}
&=& 3-2 A^{T_1}
\label{(AT1)2}  \\
(A^{T_1})^2
&=& 3+2 A^{T_1}.
\end{eqnarray}
Notice that $(A^2)^{T_1} \neq (A^{T_1})^2$, 
which is a particular case of the statement in Eq.~(\ref{A2 nge 0 4}).
The expectation values under partial transposition are          
\begin{eqnarray}
&& \langle \hat{A} \rangle_{\hat{\rho}^{T_1}}
= \langle \hat{A}^{T_1} \rangle_{\hat{\rho}}
=3,
\label{<A>rhoT1} \\
&&\langle \hat{A}^2 \rangle_{\hat{\rho}^{T1}}
= \langle [\hat{A}^2]^{T1} \rangle_{\hat{\rho}}
= -3 ,
\label{<A2>rhoT1} \\
&& \langle  (\hat{A}^{T_1})^2  \rangle_{\hat{\rho}}
= 9,
\label{<AT12>}  \\
&& {\mathrm{Var}}(\hat{A})_{\hat{\rho}^{T_1}} 
= \left\langle \left(\hat{A}-\langle \hat{A} \rangle_{\hat{\rho}^{T_1}}\right)^2\right\rangle
_{\hat{\rho}^{T_1}} 
= \langle  \hat{A}^2  \rangle_{\hat{\rho}^{T_1}} - \langle  \hat{A}  \rangle_{\hat{\rho}^{T_1}}^2
\nonumber \\
&& 
= \left\langle \left[\left(\hat{A}-\langle \hat{A}^{T_1} \rangle_{\hat{\rho}}\right)^2\right]^{T_1}\right\rangle_{\hat{\rho}}
= \left\langle  (\hat{A}^2)^{T_1}  \right\rangle_{\hat{\rho}} - \langle  \hat{A}^{T_1}  \rangle_{\hat{\rho}}^2 
\nonumber \\
&&= -12.
\label{varT1Arho}
\end{eqnarray}
Result (\ref{<AT12>}) agrees with the statement of Eq.~(\ref{AT12 ge 0}), and result
(\ref{<A2>rhoT1}) with Eq.~(\ref{A2 nge 0 5}).
From Eq.~(\ref{<A2>rhoT1}), as well as Eqs.
(\ref{varT1Arho}), the statement following Eq.~(\ref{A2 nge 0 5}) implies that the state $\hat{\rho}$ of Eq.~(\ref{rho_Phi+}) is entangled, which is indeed the case.

\section{Consequences of Theorem \ref{theorem} for a positive-definite operator expressed in terms of an arbitrary operator $\hat{\Omega}$}
\label{consequences for positive operator-Omega arbitrary}

Consider an operator $\Omega$, that is not necessarily Hermitian.
The operator $\hat{\Omega} {\hat{\Omega}}^{\dagger}$ is Hermitian and positive-definite with respect to the true state
$\hat{\rho}$, i.e.,
\begin{equation}
\langle \hat{\Omega} {\hat{\Omega}}^{\dagger} \rangle_{\hat{\rho}} 
= {\mathrm{Tr}}(\rho  \hat{\Omega} {\hat{\Omega}}^{\dagger}) \ge 0.
\label{A2 ge 0}
\end{equation}

However, $\hat{\Omega} {\hat{\Omega}}^{\dagger}$ 
may not be positive-definite with respect to the PT state
${\rho}^{T_1}$ which, in general, is not a true state and may have negative eigenvalues, i.e.,
\begin{equation}
\langle \hat{\Omega} {\hat{\Omega}}^{\dagger} \rangle_{\hat{\rho}^{T_1}}
= {\mathrm{Tr}}(\rho^{T_1} \hat{\Omega} {\hat{\Omega}}^{\dagger}) \;\;\;\;\; 
{\rm may \; not \; be} \ge 0 .
\label{A2 nge 0 11}      
\end{equation}
Theorem \ref{theorem}, in conjunction with Eq.~(\ref{A2 nge 0 11}), gives 
\begin{equation}
\langle \hat{\Omega} {\hat{\Omega}}^{\dagger} \rangle_{\hat{\rho}^{T_1}}
=\langle (\hat{\Omega} {\hat{\Omega}}^{\dagger})^{T_1} \rangle_{\hat{\rho}}
\;\;\; {\rm may \; not \; be} \ge 0 .
\label{A2 nge 0 12}
\end{equation}
The reason why the RHS of Eq.~(\ref{A2 nge 0 12}) may not be $\ge 0$, 
despite the fact that $\hat{\rho}$ is a bona fide QM state,
is that $(\hat{\Omega} {\hat{\Omega}}^{\dagger})^{T_1}$ {\em may not be a positive-definite operator}, i.e.,
\begin{equation}
\langle (\hat{\Omega} {\hat{\Omega}}^{\dagger})^{T_1} \rangle_{\hat{\rho}}
= {\mathrm{Tr}}[\rho (\hat{\Omega} {\hat{\Omega}}^{\dagger})^{T_1}] 
\stackrel{in \; general}{\neq}
{\mathrm{Tr}}[\rho \;    
\hat{\Omega}^{T_1} (\hat{\Omega}^{T_1})^{\dagger}  ] \ge 0 \; , 
\label{A2 nge 0 13}
\end{equation}
since 
\begin{equation}
(\hat{\Omega} {\hat{\Omega}}^{\dagger})^{T_1} 
\stackrel{in \; general}{\neq}
\hat{\Omega}^{T_1} (\hat{\Omega}^{T_1})^{\dagger}  \ge 0 \; . 
\label{A2 nge 0 14}
\end{equation}
In other words, 
\begin{eqnarray}
{\rm although} \;\;\; 
&& \hat{\Omega}^{T_1} (\hat{\Omega}^{T_1})^{\dagger} \ge 0 \;\; {\rm with \; respect \; to \; a \; bona \; fide \; \hat{\rho}},   
\label{AT12 ge 10}      \\
&& (\hat{\Omega} {\hat{\Omega}}^{\dagger})^{T_1} 
\stackrel{may \; not \; be}{\ge 0} \; {\rm with \; respect \; to \; a \; bona \; fide \; \hat{\rho}} .
\label{A2 nge 0 15}
\end{eqnarray}

As a result, if we find an operator $\hat{\Omega}$
such that the PT operator $(\hat{\Omega} {\hat{\Omega}}^{\dagger})^{T_1}$  
(a Hermitian operator, and thus an observable) 
has a negative expectation value in the state $\hat{\rho}$,
i.e.,
$\langle (\hat{\Omega} {\hat{\Omega}}^{\dagger})^{T_1}\rangle_{\hat{\rho}} < 0$, 
then we conclude that $\hat{\rho}$ is entangled.

If the operator $\hat{\Omega}$ is not Hermitian, it does not qualify as an observable.
However, two Hermitian operators, $\hat{H}$ and $\hat{K}$, can always be constructed
from $\hat{\Omega}$:
\begin{eqnarray}
\hat{\Omega} 
&=& \hat{H} + i \hat{K}, 
\label{Omega=H+iK}   \\
{\rm where} \;\;\; 
\hat{H}&=&  \frac12 (\hat{\Omega} + \hat{\Omega}^{\dagger}),
\;\;\; \hat{K}= \frac{1}{2i} (\hat{\Omega} - \hat{\Omega}^{\dagger}).
\label{H,K}
\end{eqnarray}
Thus, $\langle \hat{\Omega} \rangle_{\hat{\rho}}$ can be determined
by measuring the expectation value of Hermitian operators as
\begin{equation}
\langle \hat{\Omega} \rangle_{\hat{\rho}}
= \langle  \hat{H} \rangle_{\hat{\rho}}
+ i \langle  \hat{K} \rangle_{\hat{\rho}}.
\end{equation}
From Theorem \ref{theorem} we can also ``measure" 
$\langle \hat{H} \rangle_{\hat{\rho}^{T_1}}$ and
$\langle \hat{K} \rangle_{\hat{\rho}^{T_1}}$, 
and hence $\langle \hat{\Omega} \rangle_{\hat{\rho}^{T_1}}$, i.e.,
\begin{eqnarray}
\langle \hat{\Omega} \rangle_{\hat{\rho}^{T_1}}
= \langle \hat{H} \rangle_{\hat{\rho}^{T_1}}
+i \langle \hat{K} \rangle_{\hat{\rho}^{T_1}}
\\
= \langle \hat{H}^{T_1} \rangle_{\hat{\rho}}
+i \langle \hat{K}^{T_1} \rangle_{\hat{\rho}}.
\end{eqnarray}
The variance of 
$\hat{\Omega}$ in the state $\hat{\rho}^{T_1}$, i.e.,
\begin{eqnarray}
{\mathrm{Var}}(\hat{\Omega})_{\hat{\rho}^{T_1}} 
&=& \left\langle 
\left(\hat{\Omega} - \langle \hat{\Omega}\rangle_{\hat{\rho}^{T_1}}\right)
\left(\hat{\Omega} -\langle\hat{\Omega}\rangle_{\hat{\rho}^{T_1}}\right)^{\dagger}
\right\rangle_{\hat{\rho}^{T_1}} 
\label{var Omega 1} \\
&=& \left\langle  \hat{\Omega} \hat{\Omega}^{\dagger}  \right\rangle_{\hat{\rho}^{T_1}} 
-\left\langle  \hat{\Omega}  \right\rangle_{\hat{\rho}^{T_1}} 
\left\langle  \hat{\Omega}^{\dagger}  \right\rangle_{\hat{\rho}^{T_1}} 
\label{var Omega2}  \\
&=& \left\langle  \hat{\Omega} \hat{\Omega}^{\dagger}  \right\rangle_{\hat{\rho}^{T_1}} 
-\left| \left\langle  \hat{\Omega}  \right\rangle_{\hat{\rho}^{T_1}} \right|^2 \; ,
\label{var Omega2}
\end{eqnarray}
can also be expressed in terms of the observables $\hat{H}$ and $\hat{K}$.
Should this quantity be negative, the original state $\hat{\rho}$ is entangled.

\subsection{Illustration for a positive-definite operator of arbitrary dimension $N$}
\label{illustration_general_Omega}

In an $N$-dimensional Hilbert space, consider the two-particle mixed state
\begin{equation}
\hat{\rho}_r
= \frac{r}{N}
\sum_{q,q'} |qq\rangle \langle q' q'|
+ \frac{1-r}{N^2}
\sum_{q_1,q_2} |q_1 q_2\rangle \langle q_1 q_2|,
\label{isotropic-state}
\end{equation}
referred to, in the literature, as an ``isotropic state".
The pure state in the first term on the RHS is the $N$-dimension generalization of the state of Eqs.~(\ref{Phi+}), (\ref{rho_Phi+}) for $N=2$.
The second term is $1-r$ times the completely incoherent state
$\hat{\mathbb{I}}/N^2$.
The matrix elements of the isotropic state in Eq.~(\ref{isotropic-state}) 
and of its partially transposed state are
\begin{eqnarray}
\langle q_1 q_2| \hat{\rho}_r | q'_1 q'_2 \rangle
&=& \frac{r}{N}\delta_{q_1 q_2}\delta_{q'_1 q'_2}
+ \frac{1-r}{N^2}\delta_{q_1 q'_1}\delta_{q_2 q'_2}
\label{m_els_rho}  \\
\langle q_1 q_2| \hat{\rho}_r^{T_1} | q'_1 q'_2 \rangle
&=& \frac{r}{N}\delta_{q_1 q'_2}\delta_{q'_1 q_2}
+ \frac{1-r}{N^2}\delta_{q_1 q'_1}\delta_{q_2 q'_2} \; .
\label{m_els_rho_T1}
\end{eqnarray}
We shall also consider the operator
\begin{equation}
\hat{\Omega}= \sum_{m,l=0}^{N-1} x_{ml}
(\hat{X}_1^{m} \hat{Z}_1^{l})
(\hat{X}_2^{m} \hat{Z}_2^{l})^{\dagger} \; ,
\label{Omega}
\end{equation}
where $x_{ml}$ are complex coefficients.

1) The expectation value of $\hat{\Omega}$ and of $\hat{\Omega}\hat{\Omega}^{\dagger}$
in the state $\hat{\rho}_r$ are given by
\begin{eqnarray}
&& \hspace{4mm} \langle\hat{\Omega} \rangle_{\hat{\rho}_r}
= r \sum_{l=0}^{N-1} x_{0l} 
+ r\delta_{N,{\rm even}}\sum_{l=0}^{N-1} x_{\frac{N}{2},l} \;
\omega^{l\frac{N}{2}} + (1-r)x_{00} \; ,
\label{<Omega>rho} \\
&& \langle\hat{\Omega}\hat{\Omega}^{\dagger}\rangle_{\hat{\rho}_r}
= r \sum_{m,l,l'=0}^{N-1}x_{ml}x^{*}_{ml'} \; \omega^{(l-l')m} 
+(1-r)\sum_{m,l=0}^{N-1}|x_{ml}|^2  
\nonumber \\
&& \hspace{5mm} + r \; \delta_{N,{\rm even}}\sum_{m,l,l'=0}^{N-1}
\left[x_{ml}x^{*}_{m-\frac{N}{2},l'} \; \omega^{lm+(l'-2l)(m-\frac{N}{2})}
+x_{ml}x^{*}_{m+\frac{N}{2},l'} \; \omega^{lm+(l'-2l)(m+\frac{N}{2})} 
\right] \; .
\nonumber  \\
\label{<Omega><Omega+>rho} 
\end{eqnarray}
For the particular case $x_{ml}=1$ ($\forall \, m,l$), the expectation value of $\hat{\Omega}$, of $\hat{\Omega}\hat{\Omega}^{\dagger}$, and the variance of $\hat{\Omega}$ are given by
\begin{eqnarray}
&& \hspace{4mm} \langle\hat{\Omega} \rangle_{\hat{\rho}_r} 
= rN + (1-r) \; ,  
\label{<Omega> rho x=1}                    \\
&& \langle\hat{\Omega}\hat{\Omega}^{\dagger}\rangle_{\hat{\rho}_r}
=  N^2 \;,                   
\label{<Omega><Omega+> rho x=1}                   \\
&&{\mathrm{Var}}(\hat{\Omega})_{\hat{\rho}_r}
= N^2 - [rN + (1-r)]^2 \; .
\label{var Omega rho x=1}
\end{eqnarray}
One can verify that ${\mathrm{Var}}(\hat{\Omega})_{\hat{\rho}_r} \ge 0$,
$\forall \, r,N$, as it should be.

2) As outlined in Appendix \ref{proof <Omega>-var(Omega)-T1},
in the PT state $\hat{\rho}_r^{T_1}$ these expectation values are given by
\begin{eqnarray}
&& \hspace{4mm} \langle\hat{\Omega} \rangle_{\hat{\rho}_r^{T_1}}
= r\sum_{m,l=0}^{N-1}x_{ml} + (1-r)x_{00} \; ,
\label{<Omega>T1} \\
&& \langle\hat{\Omega}\hat{\Omega}^{\dagger}\rangle_{\hat{\rho}_r^{T_1}}
= r\sum_{m,l,m',l'=0}^{N-1}x_{ml} x_{m'l'}^{*} \; \omega^{ml'-m'l}
+ (1-r) \sum_{m,l=0}^{N-1}|x_{ml}|^2  \; ,
\label{<Omega><Omega>+T1} \\
&& {\mathrm{Var}}(\hat{\Omega})_{\hat{\rho}_r^{T_1}}
= r\sum_{m,l,m',l'=0}^{N-1}x_{ml} x_{m'l'}^{*} \; \omega^{ml'-m'l}
+ (1-r) \sum_{m,l=0}^{N-1}|x_{ml}|^2   
\nonumber   \\
&& \hspace{2cm} - \left|    r\sum_{m,l=0}^{N-1}x_{ml} + (1-r)x_{00}  \right|^2 ,
\label{var Omega T1}
\end{eqnarray}
where $\omega=e^{2 \pi i/N}$.  For the particular case $x_{ml}=1$ ($\forall \,  m,l$), we find
\begin{equation}
{\mathrm{Var}}(\hat{\Omega})_{\hat{\rho}_r^{T_1}}
= r\sum_{m,l,m',l' =0}^{N-1} \; \omega^{ml'-m'l}
+ (1-r)N^2 - 
\left[rN^2+(1-r)\right]^2 \; .
\label{var Omega T1 x=1}
\end{equation}
For a given $N$, this expression becomes negative for $r_0<r\leq1$, signalling entanglement of the original state. 
Using Eq.~(\ref{var Omega T1 x=1}) we determined, 
for various $N$s, the values of $r_0$ indicated in 
Table \ref{r0(N)}.  These results are consistent with $r_0=1/(N+1)$, as found, e.g., in Refs.~\cite{vidal_werner} and \cite{arunachalam_et_al2015}.

\begin{table}[ht]
\caption{For various $N$s, the value of $r_0$ for which 
${\mathrm{Var}}(\hat{\Omega})_{\hat{\rho}_r^{T_1}}$
of Eq.~(\ref{var Omega T1 x=1}) becomes negative, indicating entanglement
of the original state for $r_0<r\leq 1$. 
}
\begin{tabular}{| c
| c | 
}
\hline
$N$ & $r_0$  
\\
\hline\hline
$2$
&
$1/3$
\\
\hline
$3$
& $1/4$
\\
\hline
$4$
& $1/5$
\\
\hline
$5$
& $1/6$
\\
\hline
$9$
& $1/10$ \\
\hline
$20$
& $1/21$
\\
\hline
$50$
& $1/51$ \\
\hline\hline
\end{tabular}
\label{r0(N)}
\end{table}

\section{Summary and Conclusions} \label{conclusion}

In summary, we have shown that partial transposition for pure and mixed two-particle states in a discrete $N$-dimensional Hilbert space is equivalent to a change in sign of  a ``momentum-like" variable of one of the particles in the Wigner function for the state, thereby generalizing a result obtained for continuous-variable systems \cite{simon_2000} to the discrete-variable system case.  Therefore, the geometric interpretation of the partial transpose as a mirror reflection in phase space holds also for finite-dimensional case (although our geometric intuition is much less developed for this case).  We also showed that the expectation value of an observable in a partially transposed state can be determined via measurement, in spite of the fact that the latter may not be a physical state.  We illustrated this with the example of an isotropic state.  Hence, it is possible in principle to detect a violation of the positivity of an otherwise positive-definite operator in a partially transposed state, thereby detecting entanglement of the original state.

\bigskip

\centerline{{\bf Acknowledgments}}

PAM acknowledges support by DGAPA, under contract No. IN109014 and YBB acknowledges support from the DFG through the DIP program (FO703/2-1).

\bigskip

\begin{appendix}

\section{Schwinger operators for one particle}
\label{schwinger-one-particle}

We consider an $N$-dimensional Hilbert space spanned by $N$ distinct states $|q\rangle$, with $q=0,1, \cdots ,(N-1)$, which are subject to the periodic condition $|q+N\rangle=|q\rangle$.  These states are designated as the ``reference basis" of the space.  We follow Schwinger \cite{schwinger} and introduce the unitary operators $\hat{X}$ and $\hat{Z}$, defined by their action on the states of the reference basis by the equations
\begin{eqnarray}
\hat{Z}|q\rangle
&=&\omega^q \, |q\rangle, \;\;\;\; \omega=e^{2 \pi i/N},
\label{Z}  \\
\hat{X}|q\rangle &=& |q+1\rangle .
\label{X}
\end{eqnarray}
\label{Z,X}
The operators $\hat{X}$ and $\hat{Z}$ fulfill the periodicity condition
\begin{equation}
\hat{X}^N = \hat{Z}^N = 
\hat{\mathbb{I}},
\label{X,Z periodic}
\end{equation}
$\hat{\mathbb{I}}$ being the unit operator.
These definitions lead to the commutation relation
\begin{equation}
\hat{Z}\hat{X}=\omega \, \hat{X}\hat{Z} .
\label{comm Z,X}
\end{equation}
The two operators $\hat{Z}$ and $\hat{X}$ form a complete algebraic set, in that only a multiple of the identity commutes with both \cite{schwinger}.  As a consequence, any operator defined in our $N$-dimensional Hilbert space can be written as a function of $\hat{Z}$ and $\hat{X}$.  We also introduce (i.e., define) the Hermitian operators  $\hat{p}$ and $\hat{q}$, which play the role of ``momentum" and ``position", through the equations \cite{de_la_torre-goyeneche,durt_et_al}
\begin{eqnarray}
\hat{X}
&=& \omega^{-\hat{p}}
= e^{-\frac{2\pi i}{N}\hat{p}} \; ,
\label{X(p)}     \\
\hat{Z}
&=& \omega^{\hat{q}}
=e^{\frac{2\pi i}{N}\hat{q}} \; .
\label{Z(q)} 
\end{eqnarray}
\label{X(p),Z(q)}
What we defined as the reference basis can thus be considered as the ``position basis".  With (\ref{comm Z,X}) and definitions (\ref{X(p)}), (\ref{Z(q)}), the commutator of $\hat{q}$ and $\hat{p}$ in the continuous limit 
\cite{de_la_torre-goyeneche,durt_et_al} is the standard one, $[\hat{q},\hat{p}]=i$. 

\section{Proof of Eq.~(\ref{P(p1,p2)})}
\label{property of P(p1,p2)}

The joint probability distribution of the two momenta $p_1, p_2$ in the state $\hat{\rho}$ is given by
\begin{eqnarray}
{\cal P}_{\hat{\rho}}(p_1,p_2)
&=& {\mathrm{Tr}} (\hat{\rho} \; \mathbb{P}_{p_1} \otimes \mathbb{P}_{p_2}) 
\label{P(p1,p2) a}   \\
&=&\sum_{n_1 n_2 n_1'n_2'}
\langle n_1, n_2 | \hat{\rho} | n_1', n_2' \rangle
\langle n_1', n_2'| \mathbb{P}_{p_1} \otimes \mathbb{P}_{p_2}|n_1, n_2 \rangle   
\label{P(p1,p2) b}   \\
&=&\frac{1}{N^2}\sum_{n_1 n_2 n_1'n_2'} 
\langle n_1, n_2 | \hat{\rho} | n_1', n_2' \rangle
\omega^{p_1(n_1'-n_1)} 
\omega^{p_2(n_2'-n_2)} .
\label{P(p1,p2) c}      
\end{eqnarray}
\label{}
The joint probability distribution of the two momenta $p_1, p_2$ for the PT operator $\hat{\rho}^{T_1}$ is given by
\begin{eqnarray}
{\cal P}_{\hat{\rho}^{T_1}}(p_1,p_2)
&=& {\mathrm{Tr}} (\hat{\rho}^{T_1} \mathbb{P}_{p_1} \otimes \mathbb{P}_{p_2}) 
\label{P(p1,p2) a}   \\
&=&\sum_{n_1 n_2 n_1'n_2'}
\langle n_1, n_2 | \hat{\rho}^{T_1} | n_1', n_2' \rangle
\langle n_1', n_2'| \mathbb{P}_{p_1} \otimes \mathbb{P}_{p_2}|n_1, n_2 \rangle  
\label{P(p1,p2) b}   \\
&=&\frac{1}{N^2}\sum_{n_1 n_2 n_1'n_2'} 
\langle n_1, n_2 | \hat{\rho} | n_1', n_2' \rangle
\omega^{-p_1(n_1'-n_1)} 
\omega^{p_2(n_2'-n_2)} 
\label{P(p1,p2) c}   \\
&=& {\cal P}_{\hat{\rho}}(-p_1,p_2) .
\end{eqnarray}
\label{}
This proves Eq.~(\ref{P(p1,p2)}).  The above proof applies for $N>2$, since, for $N=2$, $| p\rangle = |-p\rangle$.

For the case of only one particle, the above result reduces to that of Eq.~(\ref{P(p) 1part}).

\section{Proof of Eqs.~(\ref{PT and W discrete 1part}) and 
(\ref{PT and W discrete})}
\label{proof of PT and W discrete}

We define the Wigner function for the density operator $\hat{\rho}$ as in 
Refs.~\cite{mello_revzen_2014,mann_mello_revzen_2016,revzen_epl_2012}, as
\begin{equation} 
W_{\hat{\rho}}(q_1,q_2,p_1,p_2) = {\mathrm{Tr}}\left[\hat{\rho}(\hat{P}_{q_1 p_1}\otimes \hat{P}_{q_2 p_2})\right] \; ,
\label{WF discrete 1}
\end{equation}
where $\hat{P}_{q_i p_i}$ is the ``line operator" for particle $i$, also defined in the above references. 
Explicitly, we find
\begin{equation}  
W_{\hat{\rho}}(q_1,q_2,p_1,p_2) 
= \sum_{q_1' q_2' q_1'' q_2''} 
\langle q_1', q_2'| \hat{\rho} |q_1'', q_2'' \rangle 
\delta_{q_1''+q_1',2q_1} \omega^{p_1 (q_1''-q_1')} 
\delta_{q_2''+q_2',2q_2} \omega^{p_2 (q_2''-q_2')} .
\label{WF discrete 2}
\end{equation}
By definition, the Wigner function after ${\rm PT_1}$ is then
\begin{eqnarray}
W_{\hat{\rho}^{T_1}}(q_1,q_2,p_1,p_2)
&=& \sum_{q_1' q_2' q_1'' q_2''} 
\langle q_1'', q_2'| \hat{\rho} |q_1', q_2'' \rangle 
\delta_{q_1''+q_1',2q_1} \omega^{p_1 (q_1''-q_1')} 
\delta_{q_2''+q_2',2q_2} \omega^{p_2 (q_2''-q_2')} 
\label{WF PT discrete a}
\nonumber \\ \\
(q_1'\Leftrightarrow q_1'') &=& \sum_{q_1' q_2' q_1'' q_2''} 
\langle q_1', q_2'| \hat{\rho} |q_1'', q_2'' \rangle 
\delta_{q_1'+q_1'',2q_1} \omega^{-p_1 (q_1''-q_1')} 
\delta_{q_2''+q_2',2q_2} \omega^{p_2 (q_2''-q_2')} 
\label{WF PT discrete b}   \nonumber \\ \\
&=& W_{\hat{\rho}}(q_1,q_2,-p_1,p_2) .
\end{eqnarray}
This proves Eq.~(\ref{PT and W discrete}).

For the case of only one particle, the above result reduces to that of Eq.~(\ref{PT and W discrete 1part}).

\section{Proof of Eqs.~(\ref{<Omega>T1}), (\ref{<Omega><Omega>+T1}) and
(\ref{var Omega T1})}
\label{proof <Omega>-var(Omega)-T1}

From the properties of one-particle Schwinger operators summarized in 
Appendix \ref{schwinger-one-particle} one can prove the following identities:
\begin{eqnarray}
\frac{1}{N}{\rm Tr}\left[(\hat{X}^m\hat{Z}^l)\left(\hat{X}^{m'}\hat{Z}^{l'}\right)^{\dagger}\right]
&=&\delta_{mm'}\delta_{ll'} \; ,
\label{TrXZ(XZ)+} \\
\frac{1}{N}{\rm Tr}\left[(\hat{X}^m\hat{Z}^l)\left(X^{m'}\hat{Z}^{l'}\right)^{\dagger}
\left(\hat{X}^m\hat{Z}^l \right)^{\dagger}(\hat{X}^{m'}\hat{Z}^{l'})\right]
&=& \omega^{ml'-m'l} \; .
\label{TrTrXZ(XZ)+XZ+XZ} 
\end{eqnarray}

We write the PT of the state of Eq.~(\ref{isotropic-state}) as
\begin{eqnarray}
\hat{\rho}_r^{T_1}
&=& \frac{r}{N}
\sum_{q,q'} |q'q\rangle \langle q q'|
+ \frac{1-r}{N^2}
\sum_{q_1,q_2} |q_1 q_2\rangle \langle q_1 q_2|,
\label{isotropic-state 1a}  \\
&\equiv& r \hat{\rho}' +(1-r)\hat{\rho}^{''} \; .
\label{isotropic-state 1b} 
\end{eqnarray}

For the first moment of $\hat{\Omega}$, we then find,
\begin{eqnarray}
{\rm Tr}(\hat{\Omega} \hat{\rho}')
&=& \frac{1}{N} \sum_{q,q'm,l} x_{ml} \times
 {_1}\langle q|\hat{X}_1^m \hat{Z}_1^l |q'\rangle_1 \; \times 
 {{\color{white}\Big|}}_2 \left\langle q'\left|\left(\hat{X}_2^m \hat{Z}_2^l\right)^{\dagger} 
 \right|q \right\rangle_2
\nonumber \\ 
&=& \frac{1}{N} \sum_{q,q'm,l} x_{ml}
 \langle q|\hat{X} ^m \hat{Z}^l |q'\rangle
 \left\langle q'\left|(\hat{X}^m \hat{Z}^l)^{\dagger} \right|q\right\rangle
\nonumber \\
&=& \frac{1}{N} \sum_{m,l} x_{ml}\;
{\rm Tr}\left[(\hat{X}^m \hat{Z}^l)(\hat{X}^m \hat{Z}^l)^{\dagger}\right]
 \nonumber \\
&=&\sum_{m,l} x_{ml} 
\label{TrOmega-rho'}    \\
{\rm Tr}(\hat{\Omega} \hat{\rho}{''})  
&=& \frac{1}{N^2}\sum_{m,l} x_{ml}
{\rm Tr}\left[(\hat{X_1}^m\hat{Z_1}^l)\left(\hat{X_2}^m\hat{Z_2}^l\right)^{\dagger}\right]
\nonumber \\
&=&\sum_{m,l} x_{ml} \delta_{m0}\delta_{l0}
=x_{00} \; .
\label{TrOmega-rho''}    
\end{eqnarray}
We used the identity (\ref{TrXZ(XZ)+}) to obtain Eq.~(\ref{TrOmega-rho''}).  Equations (\ref{TrOmega-rho'}) and (\ref{TrOmega-rho''}) are used to prove Eq.~(\ref{<Omega>T1}) in the text.

For the second moment of $\hat{\Omega}$ we have
\begin{eqnarray}
&& {\rm Tr}(\hat{\Omega} \hat{\Omega}^{\dagger} \hat{\rho}')
\\
&& = \sum_{m,l,m',l'} x_{ml} x_{m'l'}^* 
\frac{1}{N}\sum_{qq'}
{{\color{white}\Big|}}_1\left\langle q \left| (\hat{X}_1^m\hat{Z}_1^l)\left(\hat{X}_1^{m'}\hat{Z}_1^{l'}\right)^{\dagger}
\right|q'\right\rangle_1 \times
{{\color{white}\Big|}}_2 \left\langle q'\left| \left(\hat{X}_2^m \hat{Z}_2^l\right)^{\dagger}(\hat{X}_2^{m'}\hat{Z}_2^{l'})\right|q\right\rangle_2
\nonumber \\
&& = \sum_{m,l,m',l'} x_{ml} x_{m'l'}^* 
\frac{1}{N}\sum_{qq'}
\left\langle q \left| (\hat{X}^m\hat{Z}^l)\left(\hat{X}^{m'}\hat{Z}^{l'}\right)^{\dagger}
\right|q'\right\rangle
\left\langle q'\left| \left(\hat{X}^m \hat{Z}^l\right)^{\dagger}(\hat{X}^{m'}\hat{Z}^{l'})\right|q\right\rangle
\nonumber \\
&& \hspace{15mm} = \sum_{m,l,m',l'} x_{ml}x_{m'l'}^* 
\frac{1}{N}{\rm Tr}\left[
(\hat{X}^m\hat{Z}^l) \left(\hat{X}^{m'}\hat{Z}^{l'}\right)^{\dagger}
\left(\hat{X}^m\hat{Z}^l \right)^{\dagger}(\hat{X}^{m'}\hat{Z}^{l'})
\right] , 
\nonumber \\
&& \hspace{15mm}=\sum_{m,l,m',l'} x_{ml} x_{m'l'}^* \;
\omega^{ml'-m'l} .
\label{TrOmegaOmega+rho'}   \\ 
&& {\rm Tr}(\hat{\Omega} \hat{\Omega}^{\dagger} \hat{\rho}'')
= \sum_{m,l,m',l'} x_{ml} x_{m'l'}^* 
\frac{1}{N}
{\rm Tr_1}\left[(\hat{X}_1^m\hat{Z}_1^l)
\left(\hat{X}_1^{m'}\hat{Z}_1^{l'}\right)^{\dagger}\right]
\frac{1}{N}
{\rm Tr}_2\left[\left(\hat{X}_2^{m}\hat{Z}_2^{l}\right)^{\dagger}(\hat{X}_2^{m'}\hat{Z}_2^{l'})
\right] 
\nonumber \\ 
&& \hspace{15mm} = \sum_{m,l,m',l'} x_{ml} x_{m'l'}^* \delta_{mm'} \delta_{ll'}
= \sum_{m,l} |x_{ml}|^2
\label{TrOmegaOmega+rho''}   
\end{eqnarray}
To obtain Eq.~(\ref{TrOmegaOmega+rho'}) we made use of the identity (\ref{TrTrXZ(XZ)+XZ+XZ}), and to obtain Eq.~(\ref{TrOmegaOmega+rho''}) we made use of the identity (\ref{TrXZ(XZ)+}).
Equations~(\ref{TrOmegaOmega+rho'}) and (\ref{TrOmegaOmega+rho''}) are used to prove Eq.~(\ref{<Omega><Omega>+T1}) in the text.

From Eqs.~(\ref{<Omega><Omega>+T1}) and (\ref{<Omega>T1}) we find Eq.~(\ref{var Omega T1}) for the variance.

\end{appendix}


\end{document}